\documentclass[a4paper]{article}
\pdfoutput=1
\usepackage[utf8]{inputenc}
\usepackage[T1]{fontenc}
\usepackage{amsmath}
\usepackage{graphicx}
\usepackage{subfig}

\usepackage{csquotes}
\usepackage{backnaur}
\usepackage{doi}
\usepackage{array}
\usepackage{tikz}

\usepackage[%
  natbibapa
]{apacite}

\usepackage{listings}
\newcommand{\stick}{\rule[-.2ex]{.12em}{1.8ex}}

\title{Hacking with God: a Common Programming Language of Robopsychology and Robophilosophy}

\author{Norbert B\'atfai\thanks{batfai.norbert@inf.unideb.hu}\\Department of Information Technology\\University of Debrecen, Hungary}

\begin{document}
\maketitle
\begin{abstract}
This note is a sketch of how the concept of robopsychology and robophilosophy could be reinterpreted and repositioned in the spirit of the original vocation of psychology and philosophy. The notion of the robopsychology as a fictional science and a fictional occupation was introduced by Asimov in the middle of the last century. The robophilosophy, on the other hand, is only a few years old today. But at this moment, none of these new emerging disciplines focus on the fundamental and overall issues of the development of artificial general intelligence. Instead, they focus only on issues that, although are extremely important, play a complementary role, such as moral or ethical ones, rather than the big questions of life. We try to outline a conception in which the robophilosophy and robopsychology will be able to play a similar leading rule in the progress of artificial intelligence than the philosophy and psychology have done in the progress of human intelligence. To facilitate this, we outline the idea of a visual artificial language and interactive theorem prover-based computer application called Prime Convo Assistant. The question to be decided in the future is whether we can develop such an application. And if so, can we build a computer game on it, or even an esport game? It may be an interesting question in order for this game will be able to transform human thinking on the widest possible social scale and will be able to develop a standard mathematical logic-based communication channel between human and machine intelligence.
\end{abstract}

{\bf Keywords}: bicameral mind, artificial general intelligence, robophilosophy, robopsychology, pasigraphy rhapsody, interactive theorem prover, esport.

\section{Introduction}
At first, philosophy and then, over time, psychology have played a fundamental role in better self-knowing of ourselves and shaping human intelligence. The artificial intelligence revolution unfolding today, based on the neural and the dataflow paradigms, the big data and the deep learning, provides a historic opportunity for the emerging robophilosophy and robopsychology to play an equally fundamental leading role in further self-knowing of ourselves and shaping the machine intelligence as their mother sciences have done. The robopsychology as a fictional science was introduced by Asimov more than half a century ago \cite{Asimov}. In contrast, the notion of robophilosophy is barely a few years old \cite{Tzafestas}, \cite{Seibt}. We believe that these disciplines and professions are becoming real ones. But by our extreme approach, robophilosophy and robopsychology should have the same goals that philosophy and psychology had previously fulfilled, but now the object of research should no longer be humans but machines. We should not be content with they focus on issues that, although are important, but play only a secondary role, such as the human-robot interaction issues and moral or ethical ones. In the spirit of this, we present a comprehensive vision that demonstrates the potential role of robopsychology and robophilosophy and the relationship between human and machine intelligence.

\section{A \enquote{Westworld}-ian Inner Voice}
According to Julian Jaynes's bicameral mind theory \cite{Jaynes1}, the people who lived before Solon were schizophrenic in our present terminology\footnote{See also Mariano Sigman's Ted Talk entitled Your words may predict your future mental health, \url{https://www.ted.com/talks/mariano_sigman_your_words_may_predict_your_future_mental_health}.} \cite{Sigman}. Because they were unaware that their cognitive decisions are made by themselves. But Solon had already known that the person who is making the decisions and one who is interpreting them are really one and the same person, he is himself. Solon was probably already thinking the same way we are thinking today \cite{Jaynes2}. We know that the Jaynesian inner voice that tells us the solutions of our cognitive problems is not the voice of Gods but ours. And in our thinking, we experience every day that the communication protocol between us and our \enquote{Jaynesian Gods} is based on the spoken natural language \cite[pp. 138]{Jaynes2}.

The spoken language is a basic part of the Donaldian mythic culture\cite{Donald}. It is considered to be a very sophisticated mental system that is unique to man as the top of the human intellect. We don’t think so, because, for example, by the age of 4 everyone had already been speaking well in his/her mother tongue. In contrast, it is only after the age of 16 when the mathematical thinking factor may develop \cite{Geary}. Staying with the math theme, let's consider Cantor's notion of set. The language of naive set theory is based on the written form of spoken language on the one hand, and on the further basic elements of the Donaldian theoretical culture, on the other hand. We believe it is a pinnacle of the mythic culture but let's remember that Russell's antinomy has presented that it is not enough precise and powerful for further developments.

Jaynes's theory is almost 50 years old. If not from elsewhere, Jaynes's theory may be familiar from the sci-fi TV series Westworld\footnote{In an intuitive sense, the character of Dr. Robert Ford of season one can be considered as a robopsychologist, because he is a major developer of artificial intelligence hosts of the film, see also \url{https://www.imdb.com/title/tt0475784/}.} where in the season one episode three refers this theory as the theoretical background of programming of the hosts. Finally, regarding the acceptance of Jaynes’s theory, it should be noticed that Diuk and his colleagues (2012) \cite{Diuk} gave an impressive computer science-based evidence to support this theory. 

\section{Pasigraphy Rhapsody: the Written Voice of Thinking}

\paragraph{Artificial General Intelligence}
Nowadays, the Holy Grail of machine learning is the quest for Artificial General Intelligence (AGI) \cite{Hutter}. In the recent paper \cite{Robophl}, a work hypothesis was supposed that the Jaynesian inner voice is the same as the Donaldian homunculus and, in addition it will be the same as the AGI to be developed. In this sense, the AGI will be able to take over the place of Jaynesian Gods again or it will be able to form a communication channel to the Jaynesian Gods.
It should be noted that, in a sense, the taking over the place of Jaynesian Gods has already partially become a reality. Consider, for example, the AlterEgo system \cite{Kapur}, in which the AI assistant can hear the silent questions of the human and can give silent answers for those.

How do AGI agents revive and take over the role of the Jaynesian Gods? We believe this will be similar to what happened with the Internet search. Everyone had started to use the Google search engine because it found exactly what the users were looking for, yet even if the users did not know what they were looking for exactly. Similarly, AGI assistants are going to provide the best answers to users in a number of areas such as learning, training, or maintaining health.

\paragraph{The Main Question}
The question may arise, that if we used a more precise language than the spoken language could we understand more from what our homunculus is saying? Because our intuition suggests that the Jaynesian Gods may know more than they can tell us using the spoken natural language. To give the answer, we suggest to try to build a higher abstraction layer over the spoken language-based communication protocol between us and our Jaynesian Gods. Our first attempt in this direction is the Pasigraphy Rhapsody\footnote{\url{https://gitlab.com/nbatfai/pasigraphy-rhapsody}} (PaRa). It is a written only, a first-order logic-based artificial language that is intended to use through computers. For example, the reading and writing PaRa texts would be only possible with using computer programs.

Let's just focus on mathematics for a moment now. That is clearly a product of Donald’s theoretical culture. But typical mathematical proofs still have used many natural spoken language components today. In general, we simply still understand a mathematical proof built up in a spoken language even better than a formal one when we tell it to ourselves \cite[pp.  289.]{Avigad}. But over the past 100 years, the trend has clearly been to displace natural language elements from mathematical texts. As Buzzard \cite{Buzzard} makes clear, interactive theorem provers (ITPs) can bring about the same change in the typesettings and writing of mathematical texts today as the \TeX \ system, which reigned today, did in its time. The significance of TeX has been manifested in the fact that it has already handled the two cultures, the theoretic and mythic ones, separately. Its text mode allows the typesetting of the natural language elements of the manuscripts. And its math mode allows the typesetting of the visuographic mathematical elements. In contrast, in ITPs, the spoken natural language elements now play only an explanatory role for the human reader, since they are very same as comments in programming languages so they are completely ignored by further processing. That further processing typically is taking place in the Donaldian external memory of the theoretical culture, for example in computer programs through mathematical libraries. In this sense, we can say that ITP-based proofs can also come to life and live in external memory. In contrast, \TeX \  manuscripts are living only in the minds of mathematicians. Like if we compared the swimming to playing chess, where in the case of the latter, a lot more things are going on in the external memory such as the chess moves, the chessboard itself or the position of the chess pieces.

\paragraph{Esport}
Computer games and especially esport games, roughly speaking, take place entirely in the external memory. They can only come to life and live in external memory. The \enquote{language} of playing dominant esport games on the market today is so simple and universal that players of any nation can play with them almost immediately, without any training or education.
The young science of modern mathematical logic is based on precise formal languages that represents the pinnacle of Donaldian theoretical culture. Within these, the formal mathematical logic languages are living languages but only a few professionals speak them. If we would like that a lot of people will be able to speak a mathematical logic-based artificial language, we have to try to embed it into computer games or even more into some esport game. This is important because an esport game can shape players’ thinking on the widest possible social scale. In reality, of course, this does not mean simplifying the task of spreading precise thinking based on mathematical logic, but only taking a look to it from another point of view, since the quest for esport games is the Holy Grail of game development, just like the quest for AGI is the Holy Grail of machine learning.

\subsection{A Proto-Language for Pasigraphy Rhapsody}

Our distant goal is to construct sentences of a logic-based proto language from n-dimensional hypercubes. The logical basis could ensure that new hypercubes can be built from existing ones by using automated and interactive theorem provers.
Pasigraphy Rhapsody is our first attempt in this direction. Now let's take a closer look at the Pasigraphy Rhapsody formalization. We follow the introductory example of \cite{Batfai}. Let be given a natural language sentence $s = \text{\enquote{Mice hate cats}}$ in the sense that all mice hate all cats. The first step of the PaRa formalization is a classic step, to find a First-Order Logic (FOL) formalization for  $s$. It is a thought-provoking step that cannot be algorithmized. Presumably, it means that there is no classic computer program (aka Turing machine) that can produce a proper FOL formalization for all input natural language sentences.
 The formalization itself takes place using the syntax of the proto-language of Pasigraphy Rhapsody. The Backus-Naur form grammar of this proto-language originally can be found in \cite{Batfai} but it is also presented and supplemented in the following, where we use the notations of \cite{RusselNorvig} and \cite{Dragalin} books together.

\paragraph{Types}
We use countably infinite types, where a natural number denotes a type. Initially, we have no types. Specific new types appear during formalizations. For example, if $s = \text{\enquote{Mice hate cats}}$ is our first sentence to be formalized, we need to introduce the type \enquote{Animal}. This means that the number 1 denotes the \enquote{Animal} type during further formalizations. Accordingly, we need to maintain a table that records this assignment. In practice, we use numbers for the types, but from the viewpoint of theory of algorithms (that is, for easy use even with Turing machines), these numbers appeared in the grammar are given by the number of sticks \enquote{\stick}.
  
\begin{bnf}
\bnfprod{Index} {\bnfts{\stick} \bnfsp \bnfpn{Index}  \bnfor \bnfts{\stick}}\\
\bnfprod{Srt} {\bnfpn{Index}}
\end{bnf}

\paragraph{Variables}
We use countably infinite variables. The variables (and the constants, functions and predicates) are numbered and maintained similarly to the types.
\begin{bnf}
\bnfprod{Variable} {\bnfts{(}\bnfpn{Srt}\bnfts{.}\bnfpn{Index}\bnfts{)}}
\end{bnf}

\paragraph{Constants}
\begin{bnf}
\bnfprod{Cnst} {\bnfts{(C.}\bnfpn{Srt}\bnfts{.}\bnfpn{Index}\bnfts{)}}
\end{bnf}

\paragraph{Functions}
\begin{bnf}
\bnfprod{Fn} {\bnfts{(F.}\bnfpn{Index}\bnfpn{ParamList}\bnfts{)}}
\end{bnf}
\begin{bnf}
\bnfprod{ParamList} {\bnfes \bnfor \bnfts{(}\bnfpn{ParamListBuilder}\bnfts{)}}
\end{bnf}
\begin{bnf}
\bnfprod{ParamListBuilder} {\bnfpn{Term} \bnfor \bnfpn{Term}\bnfts{,} \bnfpn{ParamListBuilder}}
\end{bnf}

\paragraph{Predicates}
The formalization of $s = \text{\enquote{Mice hate cats}}$ introduces the new predicate assignments 
$\text{\enquote{Mouse}}=1$,
$\text{\enquote{Cat}}=2$ and $\text{\enquote{Hate}}=3$.

\begin{bnf}
\bnfprod{Pr} {\bnfts{(P.}\bnfpn{Index}\bnfpn{ParamList}\bnfts{)}}
\end{bnf}

\paragraph{Terms}
\begin{bnf}
\bnfprod{Term} {\bnfpn{Variable} \bnfor \bnfpn{Cnst} \bnfor \bnfpn{Fn}}
\end{bnf}

\paragraph{Atomic Formulas}
\begin{bnf}
\bnfprod{Atomic Formula} {\bnfpn{Pr}}
\end{bnf}


\paragraph{Formulas}
\begin{bnf}
	\bnfprod{Formula}{\bnfpn{Atomic Formula} \bnfor \bnfts{$\neg$}\bnfpn{Formula} } \\
	\bnfprod{Formula}{\bnfpn{Formula}\bnfpn{Connective}\bnfpn{Formula}}\\
    \bnfprod{Formula}{\bnfpn{Quantifier}\bnfpn{Variable}\bnfpn{Formula}}
\end{bnf}

\paragraph{Logical Connectives}
\begin{bnf}
\bnfprod{Connective} {\bnfts{$\wedge$} \bnfor \bnfts{$\vee$} \bnfor \bnfts{$\supset$} }
\end{bnf}

\paragraph{Quantifiers}
\begin{bnf}
\bnfprod{Quantifier} {\bnfts{$\exists$} \bnfor \bnfts{$\forall$} }
\end{bnf}

\paragraph{Mice hate cats}
Using this proto language we can formalize easily the sentence $s = \text{\enquote{Mice hate cats}}$ as follows.

\begin{equation}\label{eq:protol}
\forall \text{Animal.}x \forall \text{Animal.}y ( \text{P.Mouse}(x) \wedge \text{P.Cat}(y)  \supset \text{P.Hate}(x, y))
\end{equation}

or to be more precise

\begin{equation}
\forall (1.1) \forall (1.2) ( 
(P.1 (1.1)) \wedge (P.2 (1.2))  
\supset 
(P.3(1.1, 1.2))
)
\end{equation}

and finally with further precision, accordingly to the grammar

\begin{equation}
\forall (\stick . \stick) \forall (\stick . \stick \ \stick) ( 
(P . \stick (\stick . \stick)) \wedge (P . \stick \ \stick (\stick . \stick \ \stick))  
\supset 
(P . \stick \ \stick \ \stick(\stick . \stick, \stick . \stick \ \stick))
)
\end{equation}

but it is still not enough precise form. However, we will not specify further, because in practice we will use the notation in Equation \ref{eq:protol} and a completely different PaRa numbering system.

\subsubsection{The PaRa Numeration System}

As my Russian logic professor, Albert Drag\'alin often said in Hungarian, \enquote{\textit{Nem fontos a pontoss\'ag, csak az, hogy hogyan tudjuk el\'erni} - The accuracy is not important, what matters is to know how to achieve it}. In the spirit of this, the usage of grammar of the proto language is further simplified. We still keep the tables of newly introduced types, variables, constants, functions and predicates. However, the numbering system has been completely changed. We do not count newly introduced items separately, but in a common system as follows.

\begin{center}
\begin{tabular}{|c|c|}\hline
\centering
\textbf{Terminal letter} & \textbf{PaRa numerical code}   \\
\hline
$\stick$ & not used in the practice \\
\hline
$\exists$ & 1 \\
\hline
$\forall$ & 2 \\
\hline
$\neg$ & 3 \\
\hline
$\wedge$ & 4 \\
\hline
$\vee$ & 5 \\
\hline
$\supset$ & 6 \\\hline 
\end{tabular}
\end{center}

There are still countably infinite types, functions and predicates. In addition, within each given type, there are countably infinite constant and variable names. These still have to be numbered somehow. It is not too difficult by using the transfinite induction example of the book \cite[pp. 248]{Rozsa}, where the set of natural numbers can be represented as a union of countably infinite disjoint sets. 

\begin{itemize}
\item Let set $PaRa_\text{text}=\{7, 9, 11, 13, 15, 17, \dots \}$, that is with separating off the numerical codes (1-6) of terminal sysmbols from odd numbers we get the indices of sentences already formalized. If we hold all formalized sentences in a table, then the index of the n-th formalized sentence is $\frac{\text{x} - 1}{2}-2$ where $x$ is the n-th element of the ordered $PaRa_\text{text}$.
\item Let set $PaRa_\text{pred}=\{10, 14, 18, 22, 26, \dots \}$, that is with separating off the numerical codes of terminal sysmbols from numbers that are divisible only by the first power of 2 we get the PaRa predicate names.
\item Let set $PaRa_\text{fn}=\{12, 20, 28, 36, 44, \dots \}$, that is with separating off the numerical codes of terminal sysmbols from numbers that are divisible only by the second power of 2 we get the PaRa function names. 
\item Let set $PaRa_\text{srt}=\{8, 24, 40, 56, 72, \dots \}$, that is the numbers that are divisible only by the third power of 2 denote the PaRa type names.
\item Let set $PaRa_{c_1}=\{16, 48, 80, 112, 144, \dots \}$, that is the numbers that are divisible only by the fourth power of 2 denote the countably infinite constants of the PaRa first type.
\item Let set $PaRa_{v_1}=\{32, 96, 160, 224, 288, \dots \}$, that is the numbers that are divisible only by the fifth power of 2 denote the countably infinite variables of the PaRa first type.
\item \dots
\item Let set $PaRa_{c_n}=\{2^{2n+2}, 2^{2n+2}+2^{2n+3}, 2^{2n+2}+2*2^{2n+3},  2^{2n+2}+3*2^{2n+3}, \dots \}$ , that is the numbers that are divisible only by the $2n+2$-th power of 2 denote the countably infinite constants of the PaRa n-th type.
\item Let set $PaRa_{v_n}=\{2^{2n+3}, 2^{2n+3}+2^{2n+4}, 2^{2n+3}+2*2^{2n+4},  2^{2n+3}+3*2^{2n+4}, \dots \}$ , that is the numbers that are divisible only by the $2n+3$-th power of 2 denote the countably infinite variables of the PaRa n-th type.
\item \dots
\end{itemize}

Applying this code system to the sentence $s = \text{\enquote{Mice hate cats}}$, that is to the Equation \ref{eq:protol}, we get the following assignments.

\begin{center}
\begin{tabular}{|c|c|}\hline
\centering
\textbf{Formalized object} & \textbf{PaRa numerical code}   \\
\hline
Animal.x & 32 \\
\hline
Animal.y & 96 \\
\hline
P.Mouse & 10 \\
\hline
P.Cat & 14 \\
\hline
P.Hate & 18  \\\hline 
\end{tabular}
\end{center}

\subsubsection{The Visualization of PaRa Numeration System}

\begin{figure}[h!]
    \centering
    \subfloat[PaRa numerical codes.]{{\begin{tikzpicture}[thick,scale=.88, every node/.style={scale=.88}]
\foreach \x in {1,2,...,6}
\foreach \y in {3,...,1}
{
\draw (\x,3-\y) +(-.5,-.5) rectangle +(.5,.5);
}
\draw (1,3-1) node{$2$};
\draw (2,3-1) node{$32$};
\draw (3,3-1) node{$2$};
\draw (4,3-1) node{$96$};

\draw (2,3-2) node{$10$};
\draw (3,3-2) node{$32$};
\draw (4,3-2) node{$4$};
\draw (5,3-2) node{$14$};
\draw (6,3-2) node{$96$};

\draw (2,3-3) node{$6$};
\draw (3,3-3) node{$18$};
\draw (4,3-3) node{$32$};
\draw (5,3-3) node{$96$};
\end{tikzpicture} }}%
    \qquad
    \subfloat[SMNIST 2D Lua\LaTeX \ visual codes. The first row is generated by 4:2:1:1:0:3:2:1:0:2:0:2:1:1:0:3:3:0:2:0:1:1:0, the second one is generated by 
6:1:0:2:2:0:1:1:1:3:2:1:0:2:0:2:1:1:1:2:3:0:1:1:0:1:1:3:3:0:1:0:2:1:0 and the third one is generated by  
5:1:0:2:2:0:0:0:1:3:1:0:1:3:2:1:0:2:0:3:3:0:2:0:1:1:0.]{\label{fig:smnistb}{\includegraphics [scale=.28]{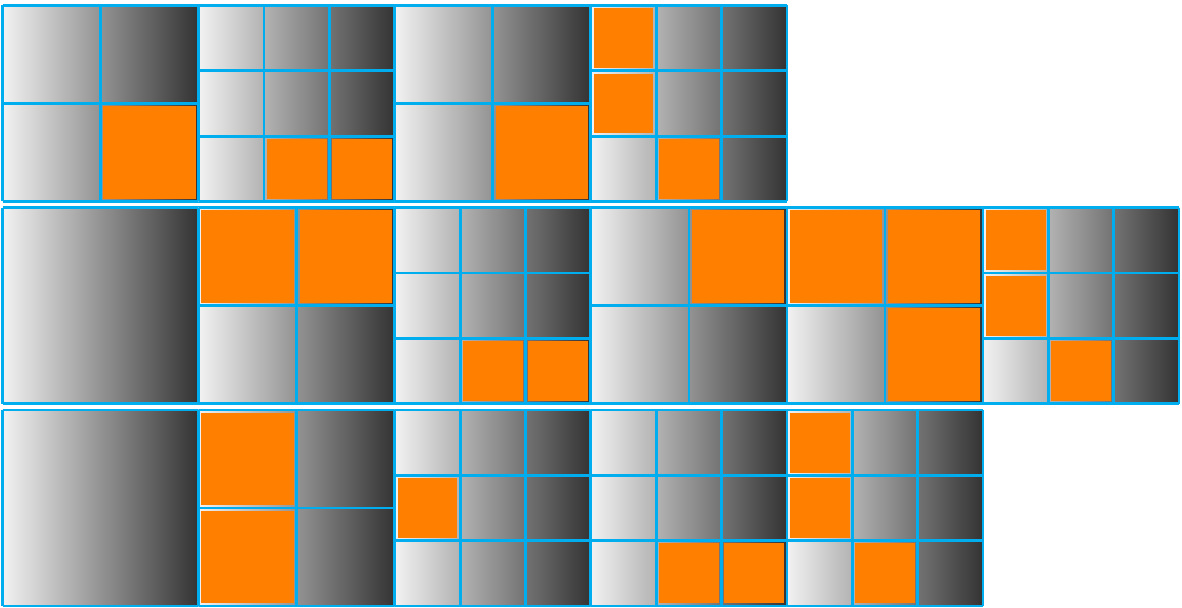}}}
    \caption{\label{fig:smnist}These figures show a 2-dimensional PaRa formalized form of the natural language sentence $s = \text{\enquote{Mice hate cats}}$.}
\end{figure}
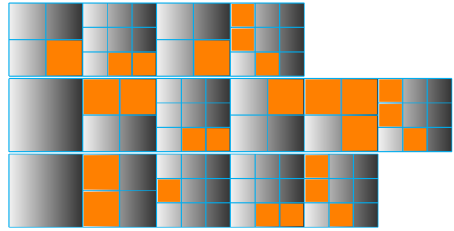

\begin{figure}[h!]
    \centering
    {\includegraphics [scale=.28]{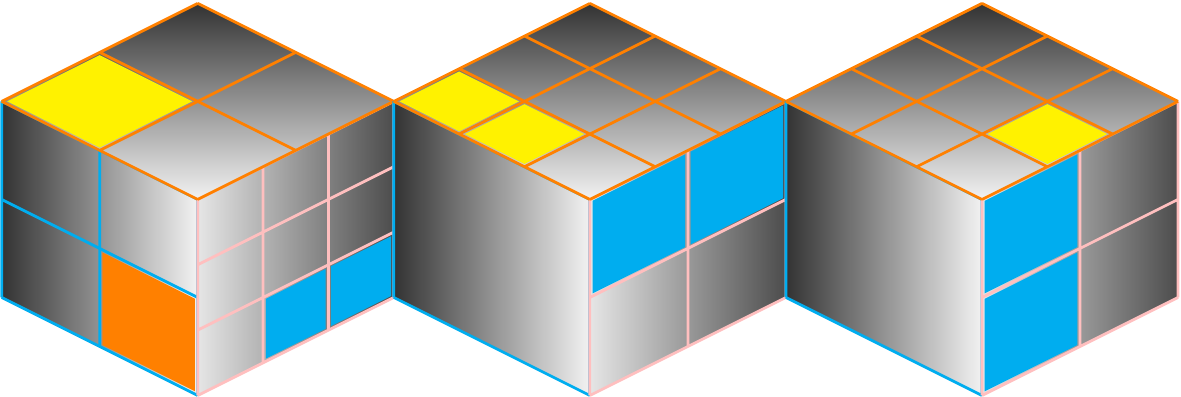}
    \caption{\label{fig:smnist3d}This figure shows a 3-dimensional PaRa formalized form of the sentence $s = \text{\enquote{Mice hate cats}}$. Only the first three 2-dimensional hypercubes of the rows of the 2-dimensional visualization are shown on a 3-dimensional hypercube. That is this visualization is generated by 3:2:1:1:0:3:2:1:0:2:0:2:1:1:0:1:0:2:2:0:1:1:1:3:2:1:0:2:0:1:0:2:2:0:0:0:1:3:1:0:1.}}
\end{figure}

\paragraph{SMNIST hypercubes} The intuitive base of visualization of PaRa proto language sentences comes from SMNIST images \cite{SMNIST1}, \cite{SMNIST2}. These images are train and test images\footnote{\url{https://gitlab.com/nbatfai/smnist/-/tree/master/Datasets/SMNIST}} for software experiments about investigating subitizing ability of machine learning programs. During the experiments the numerosity of dots drawn in the images must be recognized. However, in the case of PaRa visualization, not only the number of points matters, but also the exact position of the points. To be more precise, we enumerate all k-combinations of $l^2$ pixels, $l = 2,3,\dots$, where $l \times l$ denotes the size of the image and $k$ denotes the number of dots as it can be seen in Listing \ref{lst:smnist}. Table \ref{lst:smnist} shows some SMNIST visual codes, where the 2-dimensional hypercubes are drawn by the code snippets shown in Listing \ref{lst:prelpara2}. The 2 and 3-dimensional SMNIST hypercubes, for example shown in Fig. \ref{fig:smnistb} and \ref{fig:smnist3d}, can be written in Lua\LaTeX \ easily by using the command \texttt{prelparaIID} and  \texttt{prelparaIIID} because these are implemented in Lua\footnote{See \url{https://gitlab.com/nbatfai/pasigraphy-rhapsody/-/blob/master/para/docs/prelpara.lua}, where the idea of using yslant and xslant to achieve a 3D effect like appearance comes from Stefan Kottwitz's example \url{http://www.texample.net/tikz/examples/sudoku-3d-cube/}.}.

Finally, it should be noticed that it might be interesting to examine the use of a combinatorial number system in the future.

\begin{lstlisting}[breaklines, caption={The enumeration of relevant k-combinations of $2^2$, where p denotes the assigned PaRa numerical code and c denotes the coords of dots in the form $SMNIST(p) = \{c\}$, for example $SMNIST(10) = \{(0, 1), (1, 1)\}$. The all relevant k-combinations of $l^2$ is based on $\sum_{l=2}^{\infty}{\sum_{k=1}^{l^2-1}{\binom{l^2}{k}}}$.},captionpos=t,label={lst:smnist}]
n: 4 k: 1 p: 1, c: 1. (0,0)
n: 4 k: 1 p: 2, c: 1. (1,0) 
n: 4 k: 1 p: 3, c: 1. (0,1) 
n: 4 k: 1 p: 4, c: 1. (1,1) 
***** 
n: 4 k: 2 p: 5, c: 1. (0,0) 2. (1,0)
n: 4 k: 2 p: 6, c: 1. (0,0) 2. (0,1)
n: 4 k: 2 p: 7, c: 1. (0,0) 2. (1,1)
n: 4 k: 2 p: 8, c: 1. (1,0) 2. (0,1)
n: 4 k: 2 p: 9, c: 1. (1,0) 2. (1,1)
n: 4 k: 2 p: 10, c: 1. (0,1) 2. (1,1)
***** 
n: 4 k: 3 p: 11, c: 1. (0,0) 2. (1,0) 3. (0,1)
n: 4 k: 3 p: 12, c: 1. (0,0) 2. (1,0) 3. (1,1)
n: 4 k: 3 p: 13, c: 1. (0,0) 2. (0,1) 3. (1,1)
n: 4 k: 3 p: 14, c: 1. (1,0) 2. (0,1) 3. (1,1)
***** 
n: 9 k: 1 p: 15, c: 1. (0,0)
...
\end{lstlisting}

\begin{table}[h!]
\begin{center}
\begin{tabular}{|m{2cm} | m{2cm} | m{2cm}|}\hline
\centering
\textbf{Formalized object} & \textbf{PaRa num code} & \textbf{SMNIST code}   \\
\hline
Animal.x & 32& \includegraphics [scale=.28]{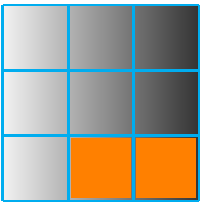}\\
\hline
Animal.y & 96&  \includegraphics [scale=.28]{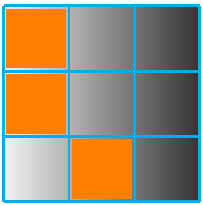}\\
\hline
P.Mouse & 10&  \includegraphics [scale=.28]{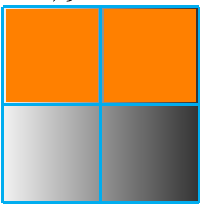}\\
\hline 
\end{tabular}
\caption{\label{tab:smnist}Some SMNIST visual codes: $SMNIST(32) = \{(1, 0), (2, 0)\}$, $SMNIST(96) = \{(1, 0), (0, 1), (0, 2)\}$ and $SMNIST(10) = \{(0, 1), (1, 1)\}$.}
\end{center}
\end{table}

\begin{lstlisting}[breaklines, caption={As this listing shows, the 2-dimensional SMNIST visual codes are drawn by the command \texttt{prelparaIID} which takes the following colon-separated arguments \texttt{the number of 2d hypercubes : the size of a 2d hypercube : the number of dots : the first coordinate of the first dot : the second coordinate of the first dot :}  $\dots$ and so on.},captionpos=t,label={lst:prelpara2},language={tex}]
$SMNIST(96) = \{(1, 0), (0, 1), (0, 2)\}$
\begin{tikzpicture}[thick,scale=2, every node/.style={scale=2}]
\prelparaIID{"1:3:3:1:0:0:1:0:2"}
\end{tikzpicture}
\end{lstlisting}

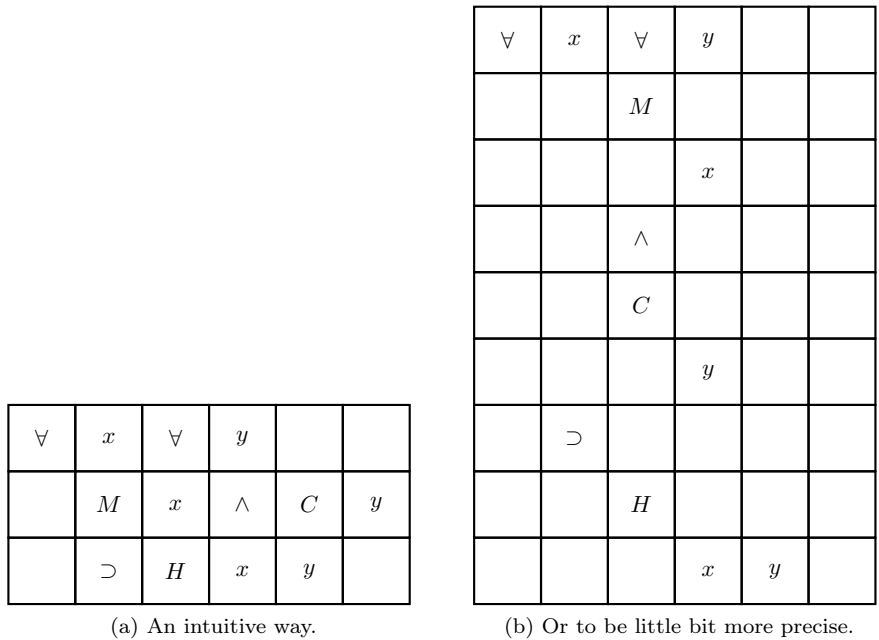
\begin{figure}[h!]
    \centering
    \subfloat[An intuitive way.]{{\begin{tikzpicture}[thick,scale=.88, every node/.style={scale=.88}]
\foreach \x in {1,2,...,6}
\foreach \y in {3,...,1}
{
\draw (\x,3-\y) +(-.5,-.5) rectangle +(.5,.5);
}
\draw (1,3-1) node{$\forall$};
\draw (2,3-1) node{$x$};
\draw (3,3-1) node{$\forall$};
\draw (4,3-1) node{$y$};

\draw (2,3-2) node{$M$};
\draw (3,3-2) node{$x$};
\draw (4,3-2) node{$\wedge$};
\draw (5,3-2) node{$C$};
\draw (6,3-2) node{$y$};

\draw (2,3-3) node{$\supset$};
\draw (3,3-3) node{$H$};
\draw (4,3-3) node{$x$};
\draw (5,3-3) node{$y$};
\end{tikzpicture} }}%
    \qquad
    \subfloat[Or to be little bit more precise.]{{\begin{tikzpicture}[thick,scale=.88, every node/.style={scale=.88}]
\foreach \x in {1,2,...,6}
\foreach \y in {9,...,1}
{
\draw (\x,9-\y) +(-.5,-.5) rectangle +(.5,.5);
}
\draw (1,9-1) node{$\forall$};
\draw (2,9-1) node{$x$};
\draw (3,9-1) node{$\forall$};
\draw (4,9-1) node{$y$};

\draw (3,9-2) node{$M$};
\draw (4,9-3) node{$x$};
\draw (3,9-4) node{$\wedge$};
\draw (3,9-5) node{$C$};
\draw (4,9-6) node{$y$};

\draw (2,9-7) node{$\supset$};
\draw (3,9-8) node{$H$};
\draw (4,9-9) node{$x$};
\draw (5,9-9) node{$y$};
\end{tikzpicture}
}}
    \caption{\label{fig:tiling}These figures show some alternatives on how to replace parentheses by indentation.}
\end{figure}

\begin{figure}[h!]
    \centering
    {\includegraphics [scale=.18]{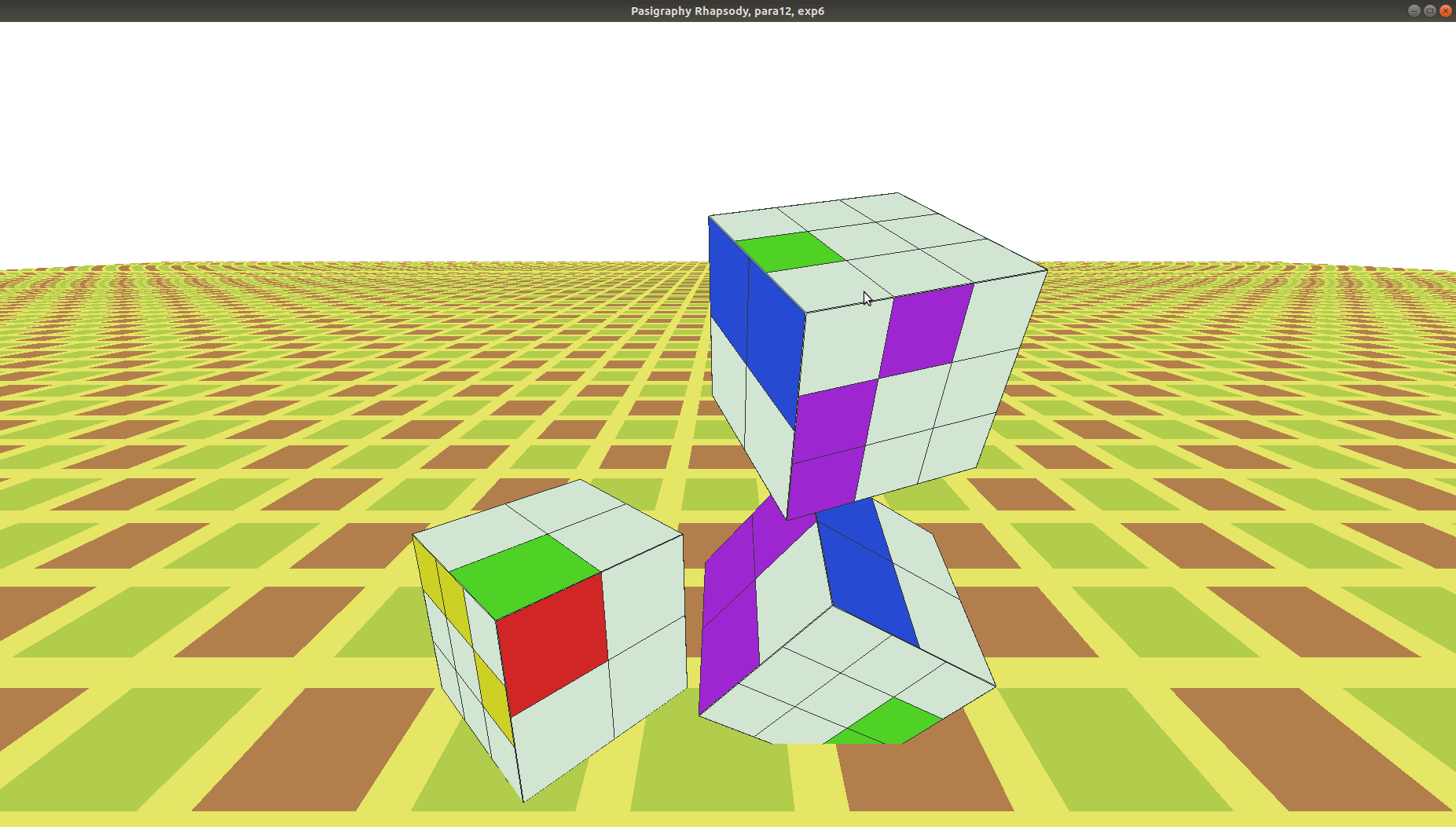}
    \caption{\label{fig:opengl}This figure shows a 3-dimensional PaRa formalized form of the sentence $s = \text{\enquote{Mice hate cats}}$ visualized in our OpenGL-based program. This visualization is generated by 3:2:1:1:0:3:2:1:0:2:0:2:1:1:0:3:3:0:2:0:1:1:0:1:0:1:0:1:0:2:2:0:1:1:1:3:2:1:0:2:0:2:1:1\-:1:2:3:0:1:1:0:1:1:3:3:0:1:0:2:1:0:1:0:2:2:0:0:0:1:3:1:0:1:3:2:1:0:2:0:3:3:0:2:0:1:1:0:1:0}}
\end{figure}

\paragraph{The PaRa Tiling Language}
In practice, parenthesizing is replaced by indentation, so using parentheses are not required. The subprocess of elimination of parentheses is referred to as PaRa tiling, in which parts of a sentence, like for example the parts of the Equation \ref{eq:protol} are placed in a tiling grid. A tiling grid for Equation \ref{eq:protol} is shown in Figure \ref{fig:tiling}. The final step in PaRa formalization is to apply SMNIST visual codes to PaRa numerical code tiles. It is shown in Figure \ref{fig:smnist} in the case of using 2-dimensional hypercubes. In the 3-dimensional case, Figure \ref{fig:smnist3d} shows a  Lua\LaTeX -based PaRa visulization of the sample sentence in question. We are also experimenting with other visualizations in the 3-dimensional case, such as our OpenGL-based renderer that can be seen in Figure \ref{fig:opengl}. In this visualization, the hypercubes can be rotated in any direction and can be completely walked around. 

\section{Robopsychology and Robophilosophy}

According to our present interpretation, it is important for both robo\-phil\-o\-sop\-hists and robo\-psycho\-lo\-gists we envision  that they should have a good understanding of formalizing natural language sentences in first-order logic. Since, on one hand, they also formalize the natural language sentences when they are working on their own PaRa implementation, and, on the other hand, the investigation of the usage of automated reasoning to produce further sentences from the knowledge that has been already gathered could be the quintessence of our research. 
Although robopsychology and robophilosophy do not focus primarily on formalizing mathematics, but it should be noticed that it would also be a seductive starting point since the formalization of whole undergraduate pure mathematics has already begun, see the paper from Kevin Buzzard titled \enquote{When will computers prove theorems?} \cite{Buzzard}. He and his colleagues are teaching mathematics not only to university students but also to machines. In this sense, our former SMNIST paper \cite{SMNIST2} investigated the teaching (to be more precise subitizing) small natural numbers to machines. But there is an important difference, that Buzzard's paper entirely works on the level of Donaldian theoretical culture, meanwhile, our paper works on the episodical culture at the abstraction level of neural networks that are usually positioned at this level \cite[pp. 363.]{Donald}. Going back to the formalization of mathematics, it could only be achievable with the help of leading experts in given specific fields. However, the PaRa translation and visualization of mathematical corpora that have already been formalized in the first-order logic would be conceivable. Nevertheless, we remain in the PaRa formalization of natural language corpora. Our preliminary idea is that the robo\-phi\-los\-o\-phists and robopsychologists would be involved on a conceptual level in the development of the business logic of PaRa software programs to be developed. 

\paragraph{Robophilosophers}
The robo\-phi\-los\-o\-phists develop the content elements of the PaRa language. It means that they introduce and maintain the new types, constants, predicates and functions of PaRa language in order to formalize natural language sentences into first-order logic (FOL) and Prolog.
A typical question for robo\-phi\-los\-o\-phists may be, for example, how to implement inheritance relationships between types. Or is it possible to create a standard library of types for formalizing natural language sentences such as the Java class library, i.e. the Java API\footnote{See \url{https://docs.oracle.com/en/java/javase/14/docs/api}. This API abstracts the world in which the programmer works during programming. For example, if the Java program must be accessed from a browser, that is it must serve requests from a browser over the HTTP protocol, the API's HTTPRequest class can be used. As its name suggests, it abstracts the HTTP request: \url{https://docs.oracle.com/en/java/javase/14/docs/api/java.net.http/java/net/http/HttpRequest.html}.} for the Java programming language? The careful selection of the elements of such a class library and the class library itself greatly can influence what and how can be expressed in the PaRa language to be developed. Another practical issue is that how to manage dictionaries for the types, constants, predicates, and functions of the PaRa language.

\paragraph{Robopsychologists} A typical question for robopsychologists may be, for example, how could PaRa visualization programs, like the mentioned Lua\LaTeX \  and OpenGL-based ones, be further developed? What new ones could be imagined in light of the latest findings, such as building on the AlterEgo system developed by Kapur et al. \cite{Kapur}. More specifically, in a PaRa-based game to be developed, where one of the player’s task may be to formalize in first-order logic of his/her everyday life events, could it be conceivable that a silent AI assistant would help players in this formalization? And in general, for this case, how should a silent AI assistant and a player communication be organized. And finally, quite generally: can an artificial visual world be built on the formalized spoken language? For example, can a Minecraft-like, \enquote{embodied} world be constructed from 3-dimensional cubes? In which the buildings corresponding to new sentences that are derived from the previous ones can be interpreted somehow?

\paragraph{Hungarian National Artificial General Intelligence}
The Hungarian Society of Robophilosophy and the Hungarian Society of Robopsychology will be two completely informally organized groups (on Facebook) with the only intended mission to create a Hungarian language reference implementation of the Pasigraphy Rhapsody which could be referred to as Hungarian National Artificial General Intelligence.
It is quite simply a matter of creating a national, in our case Hungarian, reference implementation of PaRa language by formalizing Hungarian natural language sentences. In this sense, the usage of the word \enquote{National} in the project title may be justified. According to our present interpretation, the Hungarian National Artificial General Intelligence will simply be the PaRa implementation that the Hungarian robophilosophers create together. We believe that the first successful, widespread national PaRa implementation will be able to define the mother tongue of AGI.

\section{Hacking with God: Prime Convo Assistant - pc4557}

The program to be developed, which would integrate the software components mentioned so far, is called Prime Convo Assistant (PC Asst, or in leet speak PC 4557) that would be used by both robopsychologists and robophilosophers to support them in developing National Artificial Intelligences. Using the \enquote{Prime} prefix in the name is inspired by Asimov's Prime Radiant that is a device of psychohistorians to allow writing in the language of the fictional mathematical science called Psychohistory by Asimov \cite{Asimov2}.

PC Asst must provide the following functionalities.

\begin{enumerate}
\item It must support adding new PaRa sentences and deleting obsolete ones.
\item It must support creating new PaRa dictionary elements (like types, constants, predicates and functions).
\item It must support importing and exporting PaRa dictionaries from/to PaRa dictionaries of different users.
\item It must support translating between PaRa implementations of different users (or different nations).
\item It must support yielding and/or proving sentences from formerly entered sentences using automated or interactive reasoning.
\item It must support generating sentences from formerly entered sentences using machine learning methods. 
\item It must support generating sentences from formerly entered sentences using a game interface (for example, to build a sentence from n-dimensional hypercubes like a wall of bricks). 
\end{enumerate}
The following paragraphs respond briefly to functional requirements taken in these points. 
Points 1-4 of the functional requirements for the pc4557 software to be developed are clearly related to the PaRa formalization. 
Point 5 intuitively is based on purely mathematical logic.
It is clear that formalizing the whole world seems like a completely impossible mission. That is why we are basically thinking about developing only a game for that purpose. A game, in which players formalize their own environment using their own PaRa language created for this goal. The other reason for using a game, namely, reaching the widest possible social scale, has already been discussed previously. But even to start experimenting with possible gaming experiences, we will need experience gathered with some rapid prototypes. In order to fulfill the requirement of Point 5, it would be worthwhile to try to base these prototypes on an ITP such as Lean\footnote{\url{https://github.com/leanprover}}. In parallel with the PaRa formalization, the sentences could also be formalized in Lean. Let's see for example the solution of the Barber Paradox. 

$$\neg \exists x (\text{Man}(x) \wedge \forall y ( \text{Man}(y) \supset (\text{Shaves}(x, y)  \equiv \neg \text{Shaves}(y, y)))$$

That is there is no man (a barber $x$) such that for every man $y$, $y$ is shaved by this barber $x$ if, and only if, the man $y$ does not shave himself. 
What can we do with logical sentences like this? We may try to prove it or use it in proofs of other sentences. Or adding it to a set of sentences, we may try to determine the logical consequences of this set of sentences.
For example, using \enquote{Natural deduction proof editor and checker} at \url{https://proofs.openlogicproject.org/}, Listing \ref{lst:barber1} shows\footnote{Some completely similar proofs can be found in Jon Linusson, The Barber Paradox, \url{https://gul.gu.se/public/pp/public_noticeboard_attachment/fetch?messageId=736874&fileId=18903844}. The proof contained in this paper is partially based on it.} that $\exists x (\text{Man}(x) \wedge \forall y ( \text{Man}(y) \supset (\text{Shaves}(x, y)  \equiv \neg \text{Shaves}(y, y)))$ implies a contradiction. That is, we can see for ourselves that such a barber does not exist\cite[pp. 185.]{Dragalin}.

\begin{lstlisting}[breaklines, texcl=true, literate={→}{{$\to$}}1
{¬}{{$\neg$}}1
{∃}{{$\exists$}}1
{∀}{{$\forall$}}1
{∧}{{$\wedge$}}1
{⊥}{{$\bot$}}1
{↔}{{$\leftrightarrow$}}1
, caption={This listing shows, in natural deduction, that $\exists x (\text{Man}(x) \wedge \forall y ( \text{Man}(y) \supset (\text{Shaves}(x, y)  \equiv \neg \text{Shaves}(y, y)))$ implies a contradiction. Ergo, we can see the well-known fact that such a barber does not exist.},captionpos=t,label={lst:barber1}]
1		∃x(Mx ∧ ∀y(My → (Sxy ↔ ¬Syy)))		
2			Mb ∧ ∀y(My → (Sby ↔ ¬Syy))		
3			Mb	∧E 2	
4			∀y(My → (Sby ↔ ¬Syy))	∧E 2	
5			Mb → (Sbb ↔ ¬Sbb)	∀E 4	
6			(Sbb ↔ ¬Sbb)	→E 5 3	
7				¬Sbb		
8				Sbb	↔E 6 7	
9				⊥	⊥I 7 8	
10				Sbb		
11				¬Sbb	↔E 6 10	
12				⊥	⊥I 10 11	
13			¬Sbb	¬I 10-12	
14			¬¬Sbb	¬I 7-9	
15			⊥	⊥I 13, 14	
16		⊥	∃E 1, 2-15	
17	¬∃x(Mx ∧ ∀y(My → (Sxy ↔ ¬Syy)))	¬I 1-16\end{lstlisting}

Listing \ref{lst:barber2} shows the same in Lean\footnote{The proof of the Barber paradox is an exercise of Jeremy Avigad, Robert Y. Lewis, and Floris van Doorn, Logic and Proof, \url{https://leanprover.github.io/logic_and_proof/logic_and_proof.pdf}. The proof contained in this paper is partially based on the conversations of the chatroom \url{https://leanprover.zulipchat.com/}, where in the \enquote{barber paradox} stream, the chat members have presented much nicer, shorter, more readable, human-friendly and machine-friendly proofs as well.}. It seems quite user-friendly if compared with the natural deduction. Furthermore, in Lean, we can create higher levels of abstraction, such as libraries. Accordingly, we plan to gather the initial experience required for the first rapid prototype for designing PC4557 by formalizing well-known classical AI problems (e.g., monkey and banana problems) in Lean as a part of a \enquote{Lean PaRa library}.

\begin{lstlisting}[breaklines, texcl=true, literate={→}{{$\to$}}1
{¬}{{$\neg$}}1
{∃}{{$\exists$}}1
{∀}{{$\forall$}}1
{↔}{{$\leftrightarrow$}}1
, caption={This listing shows, in Lean, that $\exists x (\text{Man}(x) \wedge \forall y ( \text{Man}(y) \supset (\text{Shaves}(x, y)  \equiv \neg \text{Shaves}(y, y)))$ implies a contradiction.},captionpos=t,label={lst:barber2}]
variables (Man : Type) (shaves : Man → Man → Prop)
theorem NoSuchBarber 
  (h : ∃ x : Man,  ∀ y : Man, shaves x y ↔ ¬ shaves y y ) : 
  false :=
  exists.elim h
  (assume barber,
    begin
      intro h,
      have hbarber : shaves barber barber ↔ ¬shaves barber barber,
        from h barber,
      have nsbb : ¬shaves barber barber, from
        assume sbb : shaves barber barber,
        show false, from hbarber.mp sbb sbb,
      show false, from nsbb (hbarber.mpr nsbb)
    end)
#print NoSuchBarber
\end{lstlisting}

Obviously, theorem proving based on natural deduction or Prolog programming cannot be the basis of the gaming experience of the game to be developed. The formalization of natural language sentences in FOL and Prolog had so far also been the default conception\footnote{See \url{https://gitlab.com/nbatfai/pasigraphy-rhapsody/blob/master/para/docs/para_prog_guide.pdf}.}. Now, this would be supplemented with a Lean-based formalization as well, initially with particular stress to the formalization of the standard AI problems, like the mentioned monkey and banana. To illustrate with an even simpler example, consider the formalization of the premises of the following well-known classic syllogism.
\begin{align*}
    \text{FOL}&:  \text{Man}(socrates), \forall x ( \text{Man}(x) \supset \text{Mortal}(x)) \\
    \text{PaRa}&:  \dots \text{drawn hypercubes} \\      
    \text{For Prolog}&:  \text{Mortal}(x) \rightleftharpoons \text{Man}(x)\\     
    \text{Prolog}&: \texttt{man(socrates). mortal(X):- man(X). }\\
    \text{Lean}&: (socrates : Man) (h: (\forall x : Man, mortal \ x)) 
 \end{align*}

Using Lean we can prove the conclusion easily. It is clear that Lean formalization does not mean a lot of extra work. And in return, Lean enables us to bring the formalized sentences to life very easily. For example, as shown in Listing \ref{lst:socrates1} and \ref{lst:socrates2}, it is easy to create a proof of the conclusion.

\begin{lstlisting}[breaklines, texcl=true, literate={→}{{$\to$}}1
{¬}{{$\neg$}}1
{∃}{{$\exists$}}1
{∀}{{$\forall$}}1
{↔}{{$\leftrightarrow$}}1
, caption={This listing shows, in Lean, a proof of the conclusion of the well-known syllogism with mortal Socrates from the premises.},captionpos=t,label={lst:socrates1}]
variables (Man : Type)  (mortal : Man → Prop)

theorem MortalSocrates (socrates : Man) (h: (∀ x : Man, mortal x)) : (mortal socrates) :=
begin
  show (mortal socrates), from h socrates,
end
\end{lstlisting}

\begin{lstlisting}[breaklines, texcl=true, literate={→}{{$\to$}}1
{¬}{{$\neg$}}1
{∃}{{$\exists$}}1
{∀}{{$\forall$}}1
{↔}{{$\leftrightarrow$}}1
, caption={This listing shows an other Lean proof of the Socrates syllogism, where the Lean automatically searches the proof.},captionpos=t,label={lst:socrates2}]
import tactic

variables (Man : Type)  (mortal : Man → Prop)

theorem MortalSocrates (socrates : Man) (h: (∀ x : Man, mortal x)) : (mortal socrates) := by tautology
\end{lstlisting}

It should be noted that within mathematics, nowadays, the formalization has received a new impetus from the motivation that the formalized databases can be used for machine learning purposes. Buzzard \cite{Buzzard} envisions that within a few years, the entire undergraduate pure mathematics will be formalized in Lean as math libraries of it. He also mentions in the same cited article that it has already begun a Lean-based project called Formal Abstracts\footnote{See \url{https://formalabstracts.github.io/} and \url{https://github.com/formalabstracts/formalabstracts}.} aimed at formalizing the theorems of articles without proofs in order to support the possibility of usage of machine learning. Narrowly interpreted, our Lean-specific goal would be very similar, but we would not focus to formalize the mathematical theorems, but formalizing the everyday life situations of standard AI problems. 

Finally, it should be noticed that there are Lean-based games, such as the Natural Number Game\footnote{\url{https://wwwf.imperial.ac.uk/~buzzard/xena/natural_number_game/}} that introduces techniques for creating interactive proofs. It is a great gamification application for teaching mathematics especially Lean. But from the point of view of gaming, it may be seen as learning mathematics rather than gaming in some classic sense. We are going to try to think of some more classical game experience, for example, a simple secret diary written in a visual or mathematical (like our PaRa) language. Where the gaming experience could be partially based on developing the player's own PaRa language itself and the visualizations of these PaRa sentences as n-dimensional structures. In AGI research, it is not really rare using big game titles, such as Minecraft, Doom, or Starcraft by the key-player companies in this industry. For example, the Microsoft uses the Minecraft MALM\"{O}\cite{MALMO}. Regardless of the PaRa language, we have already experimented with the MALM\"{O} project, it was the Red Flower Hell\footnote{\url{https://github.com/nbatfai/RedFlowerHell}} competition. In part, based on our experience with this, our intuitive impression is that the 3-dimensional PaRa hypercubes similar to those shown in Figure \ref{fig:opengl} could be placed in a Minecraft-based game world using the MALM\"{O} project. This could be a potential basis for research about the functional requirement described under Point 7.

We have not yet explained the term \enquote{Hacking with God} in the title. This term in question relates to Point 6. In research on this point, in a \enquote{Westworld}-ian sense\footnote{As it was mentioned in the cited episode of the film Westworld, Jaynes's bicameral mind idea serves as a \enquote{blueprint} for an artificial mind.}, we are going to try to design the pc4557 software architecture based on the Jaynesian idea. 

\begin{figure}[h!]
    \centering
\begin{tikzpicture}[thick,scale=.88, every node/.style={scale=.88}]
\draw [line width=1pt] (-10,5)-- (-8,5);

\draw [line width=1pt] (-8,5)-- (-8,3);
\draw [line width=1pt] (-8,3)-- (-10,3);
\draw [line width=1pt] (-10,3)-- (-10,5);
\draw [->,line width=1pt] (-9,3) -- (-9,2);
\draw [->,line width=1pt] (-9,6) -- (-9,5);
\draw [line width=1pt] (-7,7)-- (-7,1);
\draw [line width=1pt] (-6,5)-- (-4,5);
\draw [line width=1pt] (-4,5)-- (-4,3);
\draw [line width=1pt] (-4,3)-- (-6,3);
\draw [line width=1pt] (-6,3)-- (-6,5);
\draw [->,line width=1pt] (-5,6) -- (-5,5);
\draw [->,line width=1pt] (-5,3) -- (-5,2);
\draw [line width=1pt,dotted] (-3,7)-- (-3,1);
\draw [line width=1pt,dotted] (-2,5)-- (0,5);
\draw [line width=1pt,dotted] (0,5)-- (0,3);
\draw [line width=1pt,dotted] (0,3)-- (-2,3);
\draw [line width=1pt,dotted] (-2,3)-- (-2,5);
\draw [->,line width=1pt,dotted] (-1,6) -- (-1,5);
\draw [->,line width=1pt,dotted] (-1,3) -- (-1,2);
\draw [->,line width=1pt] (-6.04,0.58) -- (-8.04,0.58);
\draw (-7.,0.2) node {$\{\Gamma \to \Delta'\}$};
\draw (-7.,-.2) node {training set};

\draw (-9,6.3) node {$\Gamma$};
\draw (-5, 6.3) node {$\Gamma, S$};
\draw (-1, 6.3) node {$\Gamma, A$};

\draw (-9,1.7) node {$\Delta$};
\draw (-5, 1.7) node {$\Delta'$};
\draw (-1, 1.7) node {$\Delta''$};

\draw [->,line width=1pt] (-5.04,-1.18) -- (-9.04,-1.18);
\draw (-7.,-1.6) node {Jaynesian voice};
\draw [line width=1pt, dotted] (-9.04,-1.18) -- (-1,-1.18);

\draw (-9,4) node {Subj};
\draw (-5, 4) node {R};
\draw (-1, 4) node {TP};

\draw (-9,7.5) node {Episodic-Mythic};
\draw (-5, 7.5) node {Theoretic };
\draw (-1, 7.5) node {Esport};
\end{tikzpicture}
    \caption{\label{fig:pc4557}This is a highly speculative figure for the architectural planning of pc4557. In the top row, we indicate the appropriate Donaldian human cultural levels (from episodic to theoretic culture). The part of the figure labeled \enquote{Esport} (culture) is drawn with a dotted line because that is only our vision for the next possible level of human culture. $\Gamma$ denotes the input FOL, Prolog or Lean sentences and $\Delta$ denotes the appropriate output ones.}
\end{figure}
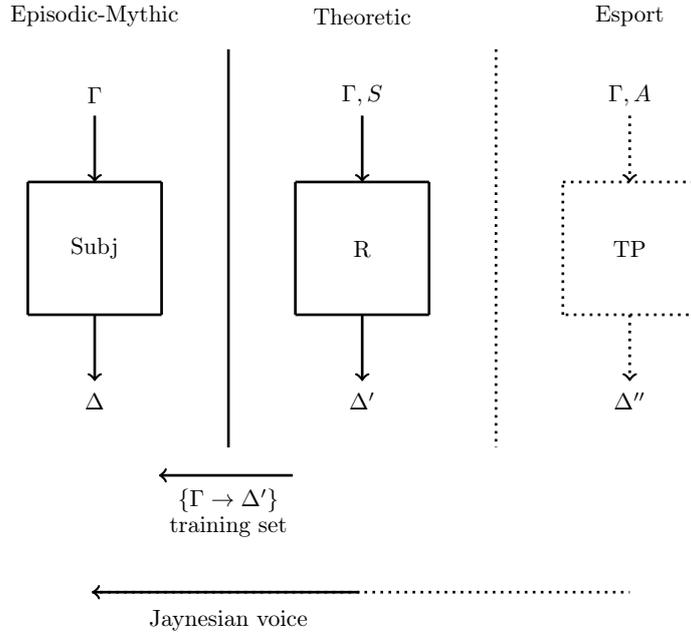

Intuitively, we feel that we have already understood something if we are not surprised on it. In a neural classification context, it means that the entropy of potential outcomes is low. This is indicated by the box labeled \enquote{Subj} in Figure \ref{fig:pc4557}. The name Subjective machine cames from paper \cite{Subj} where we assume that this putatitive subjective machine is capable of acquisition of the entropy of potential outcomes of any other neural networks (in the same way like the Turing's universal machine is capable of simulating any other Turing machines or Neumann's universal automaton is capable of reproducing itself or other automaton). It typically could be implemented by supervised learning with neural networks. 
The box labeled \enquote{R} indicates the source of the Jaynesian inner voice. For this purpose, it should have some problem-solving ability. And keep in mind that the communication protocol between the Subj part and the R part is based on a precise mathematical formal language rather than some natural language. For example, it may be a self-supervised learning component, where a BERT \cite{BERT} model could be applied to FOL, Prolog or Lean corpora as a soft theorem prover \cite{SoftTP}. 
But it should be noted that we are not aware of a BERT model for first-order logic languages.
We mentioned earlier the use of ITPs, such as the Lean, which could be incorporated into reinforcement learning models. In this context, the input labeled by $S$ denotes axioms, claims considered true or human hints. 
For example, in the very special case of predicting the next line of proofs, we could try to use the Q-learning architecture outlined in \cite{Samu}. This COP-based\footnote{Consciousness Oriented Programming, \cite{COP}.} Q-learning engine has already successfully been used by us to predict the operation of cellular automatons \cite{robom} and Turing machines\cite{trobo}. In our case, such a special solution could help to compose the $S$ sentence set.
The third part of figure outlines an even more speculative aspect, where $TP$ is not a neural network or machine learning application, but an automated theorem prover and $A$ denotes axioms that are still partially or completely unknown to mankind.
 
As a further connection between Points 5 and 6, it should be pointed out that there are results that can be considered as a symbiosis between theorem provers and machine learning. For example, \cite{Aygun} conduct experiments by teaching neural models with automatically generated theorems.

As mentioned earlier in connection with Point 7 that we could try to build a Minecraft-like game world from 3-dimensional PaRa hypercubes. For example, using our OpenGL-based visualizer, the derived new sentences must be built on a higher level than the maximum level of used premises and axioms. Our idea is that the quintessence of the Pasigraphy Rhapsody would be that there are both visual (e.g., PaRa) and textual (FOL, Prolog, or Lean) forms of the same sentence. From the point of view of potential players, using the textual elements would primarily mean learning modern mathematical logic, while the visual ones would mean the gaming. It is an open and non-trivial question whether can we equip the coveted game world, that would be isomorphic with the textual part, with rules that can give a really enjoyable gaming experience?
 
Finally, from the point of view of machine learning, we may note that the duality of textual and visual representation of the same sentences can provide the opportunity for parallel and comparative use of textual and visual machine learning algorithms.
 
\section{Conclusion}

The Pasigraphy Rhapsody presented in this paper is our first attempt to create a visual language based on mathematical logic, specifically first-order logic. While it is important to keep in mind that so far all artificial languages intended for use by humans have failed over time. Only artificial ones designed for machines, for example, the programming languages are successful \cite[pp. 38.]{Chaitin} in the long run. Therefore, it would be appropriate to try to link the Pasigraphy Rhapsody to a living and promising machine ecosystem, such as the Lean interactive theorem prover.

With our work, we would like to reduce somewhat the distance that we can discover between the worlds of programming, mathematical logic, and machine learning and the approach of the emerging robophilosophy, robopsychology, and robotheology\footnote{The concept of robo-theology can be read in context in the newspaper article titled, \enquote{Simon Nøddebo Balle – New PhD at the Department of Theology}, (Camilla Dimke, 2019), \url{https://cas.medarbejdere.au.dk/da/nyhedsbreve/nyhed/artikel/simon-noeddebo-balle-new-phd-at-the-department-of-theology/}.}.
Our poetic vision and thesis that in the esport culture to be developed as a new Donaldian stage in evolution of human mind, we will be able to hear a new Jaynesian inner voice, the voice of science, the voice of understanding. We believe that this voice will be formed and intensified on the basis of mathematical logic, and using this voice we will be able to develop the AGI and to talk to it as well.
We believe this voice was heard loud and clear by the scientists who we know by name such as Solon, Jesus Christ, Isaac Newton, Albert Einstein, or Georg Cantor.

\section{Acknowledgment}

This publication is supported by the EFOP-3.6.1-16-2016-00022 project. The project is co-financed by the European Union and the European Social Fund.

\bibliographystyle{apacite}

\bibliography{pc4557}

\begin{thebibliography}{}

\bibitem [\protect \citeauthoryear {%
Asimov%
}{%
Asimov%
}{%
{\protect \APACyear {1950}}%
}]{%
Asimov}
\APACinsertmetastar {%
Asimov}%
\begin{APACrefauthors}%
Asimov, I.%
\end{APACrefauthors}%
\unskip\
\newblock
\APACrefYear{1950}.
\newblock
\APACrefbtitle {I, Robot} {I, robot}.
\newblock
\APACaddressPublisher{}{Gnome Press}.
\PrintBackRefs{\CurrentBib}

\bibitem [\protect \citeauthoryear {%
Asimov%
}{%
Asimov%
}{%
{\protect \APACyear {1953}}%
}]{%
Asimov2}
\APACinsertmetastar {%
Asimov2}%
\begin{APACrefauthors}%
Asimov, I.%
\end{APACrefauthors}%
\unskip\
\newblock
\APACrefYear{1953}.
\newblock
\APACrefbtitle {Second foundation} {Second foundation}.
\newblock
\APACaddressPublisher{}{Doubleday}.
\PrintBackRefs{\CurrentBib}

\bibitem [\protect \citeauthoryear {%
Avigad%
}{%
Avigad%
}{%
{\protect \APACyear {2019}}%
}]{%
Avigad}
\APACinsertmetastar {%
Avigad}%
\begin{APACrefauthors}%
Avigad, J.%
\end{APACrefauthors}%
\unskip\
\newblock
\APACrefYearMonthDay{2019}{}{}.
\newblock
{\BBOQ}\APACrefatitle {Learning Logic and Proof with an Interactive Theorem
  Prover} {Learning logic and proof with an interactive theorem prover}.{\BBCQ}
\newblock
\BIn{} G.~Hanna, D.~Reid\BCBL {}\ \BBA {} M.~de Villiers\ (\BEDS),
  \APACrefbtitle {Proof Technology in Mathematics Research and Teaching} {Proof
  technology in mathematics research and teaching}\ (\BVOL~14, \BPG~277-290).
\newblock
\APACaddressPublisher{}{Springer}.
\newblock
\begin{APACrefDOI} \doi{10.1007/978-3-030-28483-1_13} \end{APACrefDOI}
\PrintBackRefs{\CurrentBib}

\bibitem [\protect \citeauthoryear {%
Aygün%
\ \protect \BOthers {.}}{%
Aygün%
\ \protect \BOthers {.}}{%
{\protect \APACyear {2020}}%
}]{%
Aygun}
\APACinsertmetastar {%
Aygun}%
\begin{APACrefauthors}%
Aygün, E.%
, Ahmed, Z.%
, Anand, A.%
, Firoiu, V.%
, Glorot, X.%
, Orseau, L.%
\BDBL {}Mourad, S.%
\end{APACrefauthors}%
\unskip\
\newblock
\APACrefYearMonthDay{2020}{}{}.
\newblock
\APACrefbtitle {Learning to Prove from Synthetic Theorems.} {Learning to prove
  from synthetic theorems.}
\newblock
\begin{APACrefURL} \url{https://arxiv.org/abs/2006.11259} \end{APACrefURL}
\PrintBackRefs{\CurrentBib}

\bibitem [\protect \citeauthoryear {%
B{\'{a}}tfai%
}{%
B{\'{a}}tfai%
}{%
{\protect \APACyear {2016}}%
}]{%
trobo}
\APACinsertmetastar {%
trobo}%
\begin{APACrefauthors}%
B{\'{a}}tfai, N.%
\end{APACrefauthors}%
\unskip\
\newblock
\APACrefYearMonthDay{2016}{}{}.
\newblock
{\BBOQ}\APACrefatitle {Theoretical Robopsychology: Samu Has Learned Turing
  Machines} {Theoretical robopsychology: Samu has learned turing
  machines}.{\BBCQ}
\newblock
\APACjournalVolNumPages{CoRR}{abs/1606.02767}{}{}.
\newblock
\begin{APACrefURL} \url{http://arxiv.org/abs/1606.02767} \end{APACrefURL}
\PrintBackRefs{\CurrentBib}

\bibitem [\protect \citeauthoryear {%
B{\'a}tfai%
}{%
B{\'a}tfai%
}{%
{\protect \APACyear {2018}}%
}]{%
Subj}
\APACinsertmetastar {%
Subj}%
\begin{APACrefauthors}%
B{\'a}tfai, N.%
\end{APACrefauthors}%
\unskip\
\newblock
\APACrefYearMonthDay{2018}{}{}.
\newblock
{\BBOQ}\APACrefatitle {{Games and artificial intelligence as the future of
  culture: an attempt to develop a theory of subjectivity}} {{Games and
  artificial intelligence as the future of culture: an attempt to develop a
  theory of subjectivity}}.{\BBCQ}
\newblock
\APACjournalVolNumPages{Inform{\'a}ci{\'o}s T{\'a}rsadalom, (only
  Hungarian)}{XVIII}{}{28--40}.
\newblock
\begin{APACrefDOI} \doi{10.3389/fnint.2012.00080} \end{APACrefDOI}
\PrintBackRefs{\CurrentBib}

\bibitem [\protect \citeauthoryear {%
B{\'a}tfai%
}{%
B{\'a}tfai%
}{%
{\protect \APACyear {2019}}%
}]{%
Batfai}
\APACinsertmetastar {%
Batfai}%
\begin{APACrefauthors}%
B{\'a}tfai, N.%
\end{APACrefauthors}%
\unskip\
\newblock
\APACrefYearMonthDay{2019}{}{}.
\newblock
\APACrefbtitle {{Construction of the Language of the Esports Culture: A
  Preliminary Study}.} {{Construction of the Language of the Esports Culture: A
  Preliminary Study}.}
\newblock
\APACrefnote{\url{https://gitlab.com/nbatfai/pasigraphy-rhapsody/-/blob/master/para/docs/con_prel_para.pdf}}
\PrintBackRefs{\CurrentBib}

\bibitem [\protect \citeauthoryear {%
B{\'a}tfai%
}{%
B{\'a}tfai%
}{%
{\protect \APACyear {2020}}%
}]{%
Robophl}
\APACinsertmetastar {%
Robophl}%
\begin{APACrefauthors}%
B{\'a}tfai, N.%
\end{APACrefauthors}%
\unskip\
\newblock
\APACrefYearMonthDay{2020}{}{}.
\newblock
{\BBOQ}\APACrefatitle {{Getting robophilosophical self-knowledge of ourselves
  in a Leibnizian language}} {{Getting robophilosophical self-knowledge of
  ourselves in a Leibnizian language}}.{\BBCQ}
\newblock
\APACjournalVolNumPages{submitted manuscript (only Hungarian)}{}{}{}.
\PrintBackRefs{\CurrentBib}

\bibitem [\protect \citeauthoryear {%
B{\'{a}}tfai%
\ \BBA {} Besenczi%
}{%
B{\'{a}}tfai%
\ \BBA {} Besenczi%
}{%
{\protect \APACyear {2017}}%
}]{%
robom}
\APACinsertmetastar {%
robom}%
\begin{APACrefauthors}%
B{\'{a}}tfai, N.%
\BCBT {}\ \BBA {} Besenczi, R.%
\end{APACrefauthors}%
\unskip\
\newblock
\APACrefYearMonthDay{2017}{}{}.
\newblock
{\BBOQ}\APACrefatitle {Robopsychology Manifesto: Samu in His Prenatal
  Development} {Robopsychology manifesto: Samu in his prenatal
  development}.{\BBCQ}
\newblock
\APACjournalVolNumPages{Carpathian Journal of Electronic and Computer
  Engineering}{10}{1}{3--12}.
\PrintBackRefs{\CurrentBib}

\bibitem [\protect \citeauthoryear {%
B{\'{a}}tfai%
\ \protect \BOthers {.}}{%
B{\'{a}}tfai%
\ \protect \BOthers {.}}{%
{\protect \APACyear {2019}}%
{\protect \APACexlab {{\protect \BCnt {1}}}}}]{%
SMNIST2}
\APACinsertmetastar {%
SMNIST2}%
\begin{APACrefauthors}%
B{\'{a}}tfai, N.%
, Papp, D.%
, Bogacsovics, G.%
, Szab{\'{o}}, M.%
, Simk{\'{o}}, V\BPBI S.%
, Bersenszki, M.%
\BDBL {}Varga, E\BPBI S.%
\end{APACrefauthors}%
\unskip\
\newblock
\APACrefYearMonthDay{2019{\protect \BCnt {1}}}{}{}.
\newblock
{\BBOQ}\APACrefatitle {Object file system software experiments about the notion
  of number in humans and machines} {Object file system software experiments
  about the notion of number in humans and machines}.{\BBCQ}
\newblock
\APACjournalVolNumPages{Cognition, Brain, Behavior. An Interdisciplinary
  Journal}{23}{4}{257--280}.
\newblock
\begin{APACrefDOI} \doi{10.24193/cbb.2019.23.15} \end{APACrefDOI}
\PrintBackRefs{\CurrentBib}

\bibitem [\protect \citeauthoryear {%
B{\'{a}}tfai%
\ \protect \BOthers {.}}{%
B{\'{a}}tfai%
\ \protect \BOthers {.}}{%
{\protect \APACyear {2019}}%
{\protect \APACexlab {{\protect \BCnt {2}}}}}]{%
SMNIST1}
\APACinsertmetastar {%
SMNIST1}%
\begin{APACrefauthors}%
B{\'{a}}tfai, N.%
, Papp, D.%
, Bogacsovics, G.%
, Szab{\'{o}}, M.%
, Simk{\'{o}}, V\BPBI S.%
, Bersenszki, M.%
\BDBL {}Varga, E\BPBI S.%
\end{APACrefauthors}%
\unskip\
\newblock
\APACrefYearMonthDay{2019{\protect \BCnt {2}}}{}{}.
\newblock
{\BBOQ}\APACrefatitle {On the notion of number in humans and machines} {On the
  notion of number in humans and machines}.{\BBCQ}
\newblock
\APACjournalVolNumPages{CoRR}{abs/1906.12213}{}{}.
\newblock
\begin{APACrefURL} \url{http://arxiv.org/abs/1906.12213} \end{APACrefURL}
\PrintBackRefs{\CurrentBib}

\bibitem [\protect \citeauthoryear {%
Buzzard%
}{%
Buzzard%
}{%
{\protect \APACyear {2020}}%
}]{%
Buzzard}
\APACinsertmetastar {%
Buzzard}%
\begin{APACrefauthors}%
Buzzard, K.%
\end{APACrefauthors}%
\unskip\
\newblock
\APACrefYearMonthDay{2020}{}{}.
\newblock
\APACrefbtitle {When will computers prove theorems?} {When will computers prove
  theorems?}
\newblock
\APACrefnote{\url{http://wwwf.imperial.ac.uk/~buzzard/xena/computers.pdf}}
\PrintBackRefs{\CurrentBib}

\bibitem [\protect \citeauthoryear {%
Bátfai%
}{%
Bátfai%
}{%
{\protect \APACyear {2011}}%
}]{%
COP}
\APACinsertmetastar {%
COP}%
\begin{APACrefauthors}%
Bátfai, N.%
\end{APACrefauthors}%
\unskip\
\newblock
\APACrefYearMonthDay{2011}{}{}.
\newblock
\APACrefbtitle {Conscious Machines and Consciousness Oriented Programming.}
  {Conscious machines and consciousness oriented programming.}
\newblock
\begin{APACrefURL} \url{https://arxiv.org/abs/1108.2865} \end{APACrefURL}
\PrintBackRefs{\CurrentBib}

\bibitem [\protect \citeauthoryear {%
Bátfai%
}{%
Bátfai%
}{%
{\protect \APACyear {2015}}%
}]{%
Samu}
\APACinsertmetastar {%
Samu}%
\begin{APACrefauthors}%
Bátfai, N.%
\end{APACrefauthors}%
\unskip\
\newblock
\APACrefYearMonthDay{2015}{}{}.
\newblock
\APACrefbtitle {A disembodied developmental robotic agent called Samu Bátfai.}
  {A disembodied developmental robotic agent called samu bátfai.}
\newblock
\begin{APACrefURL} \url{https://arxiv.org/abs/1511.02889} \end{APACrefURL}
\PrintBackRefs{\CurrentBib}

\bibitem [\protect \citeauthoryear {%
Chaitin%
}{%
Chaitin%
}{%
{\protect \APACyear {2004}}%
}]{%
Chaitin}
\APACinsertmetastar {%
Chaitin}%
\begin{APACrefauthors}%
Chaitin, G\BPBI J.%
\end{APACrefauthors}%
\unskip\
\newblock
\APACrefYearMonthDay{2004}{}{}.
\newblock
\APACrefbtitle {Meta Math! The Quest for Omega.} {Meta math! the quest for
  omega.}
\newblock
\begin{APACrefURL} \url{https://arxiv.org/abs/math/0404335} \end{APACrefURL}
\PrintBackRefs{\CurrentBib}

\bibitem [\protect \citeauthoryear {%
Clark%
, Tafjord%
\BCBL {}\ \BBA {} Richardson%
}{%
Clark%
\ \protect \BOthers {.}}{%
{\protect \APACyear {2020}}%
}]{%
SoftTP}
\APACinsertmetastar {%
SoftTP}%
\begin{APACrefauthors}%
Clark, P.%
, Tafjord, O.%
\BCBL {}\ \BBA {} Richardson, K.%
\end{APACrefauthors}%
\unskip\
\newblock
\APACrefYearMonthDay{2020}{}{}.
\newblock
\APACrefbtitle {Transformers as Soft Reasoners over Language.} {Transformers as
  soft reasoners over language.}
\PrintBackRefs{\CurrentBib}

\bibitem [\protect \citeauthoryear {%
Devlin%
, Chang%
, Lee%
\BCBL {}\ \BBA {} Toutanova%
}{%
Devlin%
\ \protect \BOthers {.}}{%
{\protect \APACyear {2018}}%
}]{%
BERT}
\APACinsertmetastar {%
BERT}%
\begin{APACrefauthors}%
Devlin, J.%
, Chang, M.%
, Lee, K.%
\BCBL {}\ \BBA {} Toutanova, K.%
\end{APACrefauthors}%
\unskip\
\newblock
\APACrefYearMonthDay{2018}{}{}.
\newblock
{\BBOQ}\APACrefatitle {{BERT:} Pre-training of Deep Bidirectional Transformers
  for Language Understanding} {{BERT:} pre-training of deep bidirectional
  transformers for language understanding}.{\BBCQ}
\newblock
\APACjournalVolNumPages{CoRR}{abs/1810.04805}{}{}.
\newblock
\begin{APACrefURL} \url{http://arxiv.org/abs/1810.04805} \end{APACrefURL}
\PrintBackRefs{\CurrentBib}

\bibitem [\protect \citeauthoryear {%
Diuk%
, Fernández~Slezak%
, Raskovsky%
, Sigman%
\BCBL {}\ \BBA {} Cecchi%
}{%
Diuk%
\ \protect \BOthers {.}}{%
{\protect \APACyear {2012}}%
}]{%
Diuk}
\APACinsertmetastar {%
Diuk}%
\begin{APACrefauthors}%
Diuk, C.%
, Fernández~Slezak, D.%
, Raskovsky, I.%
, Sigman, M.%
\BCBL {}\ \BBA {} Cecchi, G.%
\end{APACrefauthors}%
\unskip\
\newblock
\APACrefYearMonthDay{2012}{}{}.
\newblock
{\BBOQ}\APACrefatitle {A quantitative philology of introspection} {A
  quantitative philology of introspection}.{\BBCQ}
\newblock
\APACjournalVolNumPages{Frontiers in integrative neuroscience}{6}{}{}.
\newblock
\begin{APACrefDOI} \doi{10.3389/fnint.2012.00080} \end{APACrefDOI}
\PrintBackRefs{\CurrentBib}

\bibitem [\protect \citeauthoryear {%
Donald%
}{%
Donald%
}{%
{\protect \APACyear {1991}}%
}]{%
Donald}
\APACinsertmetastar {%
Donald}%
\begin{APACrefauthors}%
Donald, M.%
\end{APACrefauthors}%
\unskip\
\newblock
\APACrefYear{1991}.
\newblock
\APACrefbtitle {Origins of the modern mind: three stages in the evolution of
  culture and cognition} {Origins of the modern mind: three stages in the
  evolution of culture and cognition}.
\newblock
\APACaddressPublisher{}{Harvard University Press Cambridge, Mass}.
\PrintBackRefs{\CurrentBib}

\bibitem [\protect \citeauthoryear {%
Drag{\'a}lin%
\ \BBA {} Buz{\'a}si%
}{%
Drag{\'a}lin%
\ \BBA {} Buz{\'a}si%
}{%
{\protect \APACyear {1996}}%
}]{%
Dragalin}
\APACinsertmetastar {%
Dragalin}%
\begin{APACrefauthors}%
Drag{\'a}lin, A.%
\BCBT {}\ \BBA {} Buz{\'a}si, S.%
\end{APACrefauthors}%
\unskip\
\newblock
\APACrefYearMonthDay{1996}{}{}.
\newblock
\APACrefbtitle {Bevezetés a matematikai logikába.} {Bevezetés a matematikai
  logikába.}
\newblock
\APACaddressPublisher{}{Kossuth Egyetemi Kiadó}.
\PrintBackRefs{\CurrentBib}

\bibitem [\protect \citeauthoryear {%
Geary%
}{%
Geary%
}{%
{\protect \APACyear {1996}}%
}]{%
Geary}
\APACinsertmetastar {%
Geary}%
\begin{APACrefauthors}%
Geary, D\BPBI C.%
\end{APACrefauthors}%
\unskip\
\newblock
\APACrefYearMonthDay{1996}{}{}.
\newblock
{\BBOQ}\APACrefatitle {Biology, culture, and cross-national differences in
  mathematical ability} {Biology, culture, and cross-national differences in
  mathematical ability}.{\BBCQ}
\newblock
\BIn{} \APACrefbtitle {The nature of mathematical thinking.} {The nature of
  mathematical thinking.}\ (\BPGS\ 145--171).
\newblock
\APACaddressPublisher{}{Lawrence Erlbaum Associates, Inc}.
\PrintBackRefs{\CurrentBib}

\bibitem [\protect \citeauthoryear {%
Hutter%
}{%
Hutter%
}{%
{\protect \APACyear {2012}}%
}]{%
Hutter}
\APACinsertmetastar {%
Hutter}%
\begin{APACrefauthors}%
Hutter, M.%
\end{APACrefauthors}%
\unskip\
\newblock
\APACrefYearMonthDay{2012}{}{}.
\newblock
{\BBOQ}\APACrefatitle {One Decade of Universal Artificial Intelligence} {One
  decade of universal artificial intelligence}.{\BBCQ}
\newblock
\APACjournalVolNumPages{CoRR}{abs/1202.6153}{}{}.
\newblock
\begin{APACrefURL} \url{http://arxiv.org/abs/1202.6153} \end{APACrefURL}
\PrintBackRefs{\CurrentBib}

\bibitem [\protect \citeauthoryear {%
Jaynes%
}{%
Jaynes%
}{%
{\protect \APACyear {1976}}%
}]{%
Jaynes1}
\APACinsertmetastar {%
Jaynes1}%
\begin{APACrefauthors}%
Jaynes, J.%
\end{APACrefauthors}%
\unskip\
\newblock
\APACrefYear{1976}.
\newblock
\APACrefbtitle {The origin of consciousness in the breakdown of the bicameral
  mind} {The origin of consciousness in the breakdown of the bicameral mind}.
\newblock
\APACaddressPublisher{}{Houghton Mifflin}.
\PrintBackRefs{\CurrentBib}

\bibitem [\protect \citeauthoryear {%
Jaynes%
}{%
Jaynes%
}{%
{\protect \APACyear {1986}}%
}]{%
Jaynes2}
\APACinsertmetastar {%
Jaynes2}%
\begin{APACrefauthors}%
Jaynes, J.%
\end{APACrefauthors}%
\unskip\
\newblock
\APACrefYearMonthDay{1986}{}{}.
\newblock
{\BBOQ}\APACrefatitle {Consciousness and the voices of the mind} {Consciousness
  and the voices of the mind}.{\BBCQ}
\newblock
\APACjournalVolNumPages{Canadian Psychology}{27}{}{128-139}.
\PrintBackRefs{\CurrentBib}

\bibitem [\protect \citeauthoryear {%
Johnson%
, Hofmann%
, Hutton%
\BCBL {}\ \BBA {} Bignell%
}{%
Johnson%
\ \protect \BOthers {.}}{%
{\protect \APACyear {2016}}%
}]{%
MALMO}
\APACinsertmetastar {%
MALMO}%
\begin{APACrefauthors}%
Johnson, M.%
, Hofmann, K.%
, Hutton, T.%
\BCBL {}\ \BBA {} Bignell, D.%
\end{APACrefauthors}%
\unskip\
\newblock
\APACrefYearMonthDay{2016}{}{}.
\newblock
{\BBOQ}\APACrefatitle {The Malmo Platform for Artificial Intelligence
  Experimentation} {The malmo platform for artificial intelligence
  experimentation}.{\BBCQ}
\newblock
\BIn{} \APACrefbtitle {25th International Joint Conference on Artificial
  Intelligence (IJCAI-16).} {25th international joint conference on artificial
  intelligence (ijcai-16).}
\newblock
\APACaddressPublisher{}{AAAI - Association for the Advancement of Artificial
  Intelligence}.
\newblock
\begin{APACrefURL}
  \url{https://www.microsoft.com/en-us/research/publication/malmo-platform-artificial-intelligence-experimentation/}
  \end{APACrefURL}
\PrintBackRefs{\CurrentBib}

\bibitem [\protect \citeauthoryear {%
Kapur%
, Kapur%
\BCBL {}\ \BBA {} Maes%
}{%
Kapur%
\ \protect \BOthers {.}}{%
{\protect \APACyear {2018}}%
}]{%
Kapur}
\APACinsertmetastar {%
Kapur}%
\begin{APACrefauthors}%
Kapur, A.%
, Kapur, S.%
\BCBL {}\ \BBA {} Maes, P.%
\end{APACrefauthors}%
\unskip\
\newblock
\APACrefYearMonthDay{2018}{03}{}.
\newblock
{\BBOQ}\APACrefatitle {AlterEgo: A Personalized Wearable Silent Speech
  Interface} {Alterego: A personalized wearable silent speech
  interface}.{\BBCQ}
\newblock
\BIn{} (\BPG~43-53).
\newblock
\begin{APACrefDOI} \doi{10.1145/3172944.3172977} \end{APACrefDOI}
\PrintBackRefs{\CurrentBib}

\bibitem [\protect \citeauthoryear {%
P{\'e}ter%
}{%
P{\'e}ter%
}{%
{\protect \APACyear {1976}}%
}]{%
Rozsa}
\APACinsertmetastar {%
Rozsa}%
\begin{APACrefauthors}%
P{\'e}ter, R.%
\end{APACrefauthors}%
\unskip\
\newblock
\APACrefYear{1976}.
\newblock
\APACrefbtitle {Playing with Infinity: Mathematical Explorations and
  Excursions} {Playing with infinity: Mathematical explorations and
  excursions}.
\newblock
\APACaddressPublisher{}{Dover Publications Inc.}
\PrintBackRefs{\CurrentBib}

\bibitem [\protect \citeauthoryear {%
Russell%
\ \BBA {} Norvig%
}{%
Russell%
\ \BBA {} Norvig%
}{%
{\protect \APACyear {2010}}%
}]{%
RusselNorvig}
\APACinsertmetastar {%
RusselNorvig}%
\begin{APACrefauthors}%
Russell, S.%
\BCBT {}\ \BBA {} Norvig, P.%
\end{APACrefauthors}%
\unskip\
\newblock
\APACrefYear{2010}.
\newblock
\APACrefbtitle {Artificial Intelligence: A Modern Approach} {Artificial
  intelligence: A modern approach}\ (\PrintOrdinal{Third}\ \BEd).
\newblock
\APACaddressPublisher{}{Pearson Education}.
\PrintBackRefs{\CurrentBib}

\bibitem [\protect \citeauthoryear {%
Seibt%
}{%
Seibt%
}{%
{\protect \APACyear {2017}}%
}]{%
Seibt}
\APACinsertmetastar {%
Seibt}%
\begin{APACrefauthors}%
Seibt, J.%
\end{APACrefauthors}%
\unskip\
\newblock
\APACrefYearMonthDay{2017}{}{}.
\newblock
{\BBOQ}\APACrefatitle {Robophilosophy} {Robophilosophy}.{\BBCQ}
\newblock
\BIn{} R.~Braidotti\ \BBA {} M.~Hlavajova\ (\BEDS), \APACrefbtitle {Posthuman
  Glossary} {Posthuman glossary}\ (\BPGS\ 390--393).
\newblock
\APACaddressPublisher{}{Bloomsbury}.
\PrintBackRefs{\CurrentBib}

\bibitem [\protect \citeauthoryear {%
Sigman%
}{%
Sigman%
}{%
{\protect \APACyear {2017}}%
}]{%
Sigman}
\APACinsertmetastar {%
Sigman}%
\begin{APACrefauthors}%
Sigman, M.%
\end{APACrefauthors}%
\unskip\
\newblock
\APACrefYear{2017}.
\newblock
\APACrefbtitle {The Secret Life of the Mind: How Your Brain Thinks, Feels, and
  Decides} {The secret life of the mind: How your brain thinks, feels, and
  decides}.
\newblock
\APACaddressPublisher{}{Little, Brown Spark}.
\PrintBackRefs{\CurrentBib}

\bibitem [\protect \citeauthoryear {%
Tzafestas%
}{%
Tzafestas%
}{%
{\protect \APACyear {2016}}%
}]{%
Tzafestas}
\APACinsertmetastar {%
Tzafestas}%
\begin{APACrefauthors}%
Tzafestas, S\BPBI G.%
\end{APACrefauthors}%
\unskip\
\newblock
\APACrefYear{2016}.
\newblock
\APACrefbtitle {An Introduction to Robophilosophy Cognition,
  Intelligence,Autonomy, Consciousness, Conscience, and Ethics} {An
  introduction to robophilosophy cognition, intelligence,autonomy,
  consciousness, conscience, and ethics}.
\newblock
\APACaddressPublisher{}{River Publishers}.
\PrintBackRefs{\CurrentBib}

\end{thebibliography}

\end{document}